# On Creating a Comprehensive Food Database

Lexington Whalen
College of Computing and Engineering
University of South Carolina
Columbia, SC, United States of America
LAWHALEN@email.sc.edu

Brie Turner-McGrievy
Arnold School of Public Health
University of South Carolina
Columbia, SC, United States of America
brie@sc.edu

Matthew McGrievy
Arnold School of Public Health
University of South Carolina
Columbia, SC, United States of America
mjm@sc.edu

Andrew Hester
Arnold School of Public Health
University of South Carolina
Columbia, SC, United States of America
hesteras@mailbox.sc.edu

Homayoun Valafar
College of Computing and Engineering
University of South Carolina
Columbia, SC, United States of America
homayoun@cse.sc.edu

*Abstract*—Studies with the primary aim of addressing eating disorders focus on assessing the nutrient content of food items with an exclusive focus on caloric intake. There are two primary impediments that can be noted in these studies. The first of these relates to the fact that caloric intake of each food item is calculated from an existing database. The second concerns the scientific significance of caloric intake used as the single measure of nutrient content. By requiring an existing database, researchers are forced to find some source of a comprehensive set of food items as well as their respective nutrients. This search alone is a difficult task, and if completed often leads to the requirement of a paid API service. These services are expensive and non-customizable, taking away funding that could be aimed at other parts of the study only to give an unwieldy database that can not be modified or contributed to. In this work, we introduce a new rendition of the USDA's food database that includes both foods found in grocery stores and those found in restaurants or fast food places. At the moment, we have accumulated roughly 1.5 million food entries consisting of approximately 18,000 brands and 100 restaurants in the United States. These foods also have an abundance of nutrient data associated with them, from the caloric amount to saturated fat levels. The data is stored in MySQL format and is spread among five major tables. We have also procured images for these food entries when available, and have included all of our data and program scripts in an open source repository that anyone can access, for free.

*Keywords—MySQL, Table Design, USDA, Food Database*

## I. Introduction

The United States Department of Agriculture (USDA) provides a comprehensive database of foods for sale within the United States [1]. This database, FoodData Central, can be downloaded for free at usda.gov, and contains several sub databases. A description of each database can be found in USDA DATABASES. The USDA currently (June 2022) supplies database files as JSON and CSV. In the past, the Microsoft Access filetype was also supported, however this was discontinued in October of 2021 [2]. While it does contain useful data, it is rather convoluted and in general more complex than it needs to be. This lack of user-friendliness is a deterrent to researchers in search of a quick and easy database to obtain food or nutrient information from. More specifically, the database contains legacy conent that is no longer supported, nutritional contents reported by different vendors, and many other components that are genarally not used by researchers in non-technical domains. The proper nomalization of the additonal information adds many layers of complexity to the database schema, which also impeded their easy use. Finally, some critical relationships between the large number of tables are not clearly defined, severely reducing the effective use of the database.

While comprehensive with regards to foods for sale at a grocery store, the database also lacks many entries for common restaurant foods; this is a major flaw as it is known that roughly 40% of all adults in the United States eat fast food on any given day [3], and that 60% of Americans admit to eating dinner out at least once a week [4]. This paper offers an explanation of different sources and methods of data collection, and compares their relative merits. The original study that generated the motivation for this paper chose Menustat, a database built originally by the New York City Department of Health and Mental Hygiene [5]. More details on Menustat may be found in the section MENUSTAT DATABASE.

## II. Background

Many studies funded by international, national, and regional agencies are designed to help participants lose weight or modify their diet.

In many studies, technologies such as smart devices are used to record the nutritional intake of participants via logging of meals for each day. In order to do this, a supply of food candidates is the central database requirement of these studies. A recent example of such a study is found in SocialPOD. SocialPOD combined a social network where users could "trade pounds for points" by interacting with eachother, posting weight updates, meals, or comments [6]. SocialPOD thus required a database of foods that users could choose from, but could not find any viable free sets of data, and was forced to pay for an expensive API. Researchers from the SocialPOD project then expressed interest in the creation of an open source comprehensive food database, leading to the creation of this paper. Due to unavailability of useful and usable databases, these studies resort to the use of expensive commercially available databases such as Nutritionix [7], a popular food database API, and thus lose funds that could be more efficiently allocated to other operation costs.

### III. USDA DATABASES

The USDA provides several options for database downloads, and also provides an API to query their online database. However, due to the complexities of the data, the download and reconstitution of the databases in a usable form is a very daunting task. The API provided by USDA is relatively easy to use but it is limited in scalibity of use as the number of API requests are limited on daily basis, and the speed of serving each request is relatively slow. USDA provides the downloadable versio nof their database in the two popular database file types JSON and CSV.

As of October 2021, there are a total of 37 CSV files available in the entire downloadable content, however only a select few are relevant to calculating macronutrient amounts [1]. These files are listed in Table 1. Note that file names are denoted by <FILE_NAME>.

| File name | Size |
|---|---|
| <branded_food.csv> | 745 M |
| <food.csv> | 140 M |
| <food_nutrient.csv> | 1.3 G |
| <food_portion.csv> | 5.8 M |
| <foundation_food.csv> | 9 K |
| <nutrient.csv> | 21 K |
| <sr_legacy_food.csv> | 128 K |

*Table 1: Files and their sizes.*

Each CSV file was parsed and loaded into MySQL using the *--local-infile* flag; this option allows a file that exists on a client computer to be loaded rather than requiring the server to have the file, which saves much time. It is important to note that the loading of the files was done via the MySQL command line, as the GUI version will lead to much slower loading times.

The table definitions were generated using MySQL Workbench's Import Wizard. The Import Wizard is only necessary to run for approximately a second before a safe closure is possible since all that is needed is the table definitions; the actual values are to be imported via command line. Of course while the table definitions could also be done manually, it is much simpler to use MySQL Workbench as there are often many columns. When using the Import Wizard, it is important to keep in mind that sometimes the suggested datatypes are incorrect; for instance, fdc_id should be of type BigInt, not just int (if just int overflow will occur when adding data). The Data Import Wizard is shown in Figure 1; note that one can select from the recognized columns and change their data types.

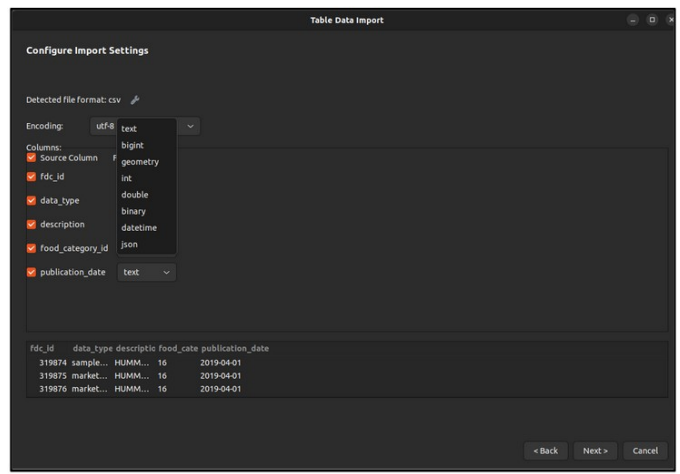

Figure 1: The Table Data Import Wizard

As an example, the following command was ran in an Ubuntu 20.04 MySQL terminal to load food_nutrient.csv, where values in brackets are to be replaced with the corresponding values from the respective machine:

```
load data local infile [PATH_TO_CSV] into table [TABLE_NAME] fields terminated by ',' enclosed by '"' lines terminated by '\n' ignore 1 rows
```

*Dialogue 1. The command used to load a csv file into the database, using the MySQL command line. This was done to make loading of larger files (such as <food_nutrient.csv>) faster.*

While the command is relatively easy to understand, some common mistakes include forgetting to set the --load-infile flag to 1 or ignoring the first row (due to the fact that the first row consists of column names only, and not values).

An important and interesting point of consideration is that this method was used due to the fact that MySQL Workbench is considerably slower than using the terminal; while MySQL Workbench would consistently take hours or days to load large files, the command line version took at most an hour for even the largest files.

Upon importing the necessary tables, the next step is to create indexes for them. An index is essentially a data structure that can be created with one or more columns and allows for increases in operation speeds.

The following MySQL command can be used to add an index for the fdc_id column of table branded_food:

```
ALTER TABLE branded_food
ADD INDEX fdc_id(fdc_id)
```
*Dialogue 2. Adding an index to a table.*

The important columns to index are fdc_id for all tables that include an fdc_id column and nutrient_id from food_nutrient. Any other columns that are indexable may also be indexed for further performance improvements, however these two are the most important as the creation of the tables that follow make great use of them.

The USDA food database can be split into four categories: *Branded Foods*, *Foundational Foods*, *SR Legacy Foods*, and *Experimental Foods*. To split the data into these four tables, the following general approach can be use:

1. Select the columns needed from food or other tables (ie: *Branded*) using SELECT
2. Join the tables on fdc_id
3. Join with food_nutrient on the respective id

As an example, in order to create a table with only entries from sr legacy containing columns fdc_id, description, kcals, servings, gram_weight one can perform the following command:

```
CREATE TABLE sr_legacy_food AS
    select fs.fdc_id, fs.description, fn.amount as kcals,
    fp.amount as servings, fp.gram_weight
from food_sr_legacy_food fs
inner join food_nutrient fn
    on fn.fdc_id = fs.fdc_id
inner join food_portion fp
    on fs.fdc_id = fp.fdc_id
where
    -- id 1008 is the id for Kilocalories
    fn.nutrient_id = 1008
```
*Dialogue 3. How the food tables can be created. The example shows how one can access food nutrients for a specific table.*

Note that for specific nutrient values, an inspection of food_nutrient is needed (see the 1008 in the above command).

## IV. MENUSTAT

*Menustat* is a food database that specializes in restaurant data. Figure 2 illustrates the database; note the inclusion of restaurant, food category, serving size, and macronutrient information. Also note that there are more macronutrients available than just energy. The data is recorded annually, and it can be downloaded as either excel or csv file types [5]. These files can be imported in a similar fashion as the USDA data, by first using the *Import Wizard* and then the command line. One common issue when importing the data is that some rows may be formatted imperfectly, leading to missing row information. This can be fixed by converting the file to a known encoding by running cat -v [NAME OF FILE].csv > [NEW NAME OF FILE].csv on a Linux machine. After running this command some end lines will now have turned into ^M; these can be removed using *Vim* or any text editor of choice.

When deciding the data types of the columns, it is easiest to assign all columns the TEXT type; this ensures that all rows will be correctly loaded, since when attempting to load data into columns of other types lead to rows being skipped as the data was too large. Upon completion of the load, these column types may be changed.

Figure 2: A snippet of the Menustat database.

## V. TABLE DESIGN

Upon creating the tables, there were two general ways to set them up based on how one desires to handle the nutrients. The first way is a "row-based" approach, where each food item has a column labeled nutrient_name, and has multiple rows describing each nutrient of the food. For example, an entry for Broccoli would have each row be a new entry for a nutrient; the only column that would be changing would be the name and value of the nutrient, and the serving size for different serving sizes. Figure 3 shows the row-based approach; note that data entered "row-wise" in that each row contains the next nutrient type. Contrast with the "column-based" approach of Figure 4.

The benefit of this approach is the ease at which new nutrient values can be added; for example, if Broccoli were to have its vitamin D measured, only a new row containing this value would have to be added; no new column necessary. The fact that no new column is necessary is advantageous can be easily seen by imagining that all nutrient categories were to be stored as column values. This would mean that if Broccoli were to add a new nutrient that has previously not been recorded, then a new column would have to be introduced for the entirety of the database rather than just adding one more row that includes this new nutrient. The main disadvantage is that this approach is slightly difficult to work with; if one knows in advance what nutrients are to be queryable, it may be easier to simply return one row with all of the possible nutrients within. This is the motivation for the second approach.

| description | brand_owner | serving_size | serving_size_un | nutrient_name | nutrient_unit | nutrient_amount |
|---|---|---|---|---|---|---|
| WESSON Vegetable Oil 1 GAL | Richardson Oilseed Products (US) … | 15 | ml | fatty acids, total… | G | 8 |
| WESSON Vegetable Oil 1 GAL | Richardson Oilseed Products (US) … | 15 | ml | fatty acids, total… | G | 3 |
| WESSON Vegetable Oil 1 GAL | Richardson Oilseed Products (US) … | 15 | ml | fatty acids, total… | G | 2 |
| WESSON Vegetable Oil 1 GAL | Richardson Oilseed Products (US) … | 15 | ml | energy | KCAL | 130.05 |
| WESSON Vegetable Oil 1 GAL | Richardson Oilseed Products (US) … | 15 | ml | total lipid (fat) | G | 14 |
| SWANSON BROTH BEEF | CAMPBELL SOUP COMPANY | 240 | ml | sugars, total inc… | G | 1.01 |
| SWANSON BROTH BEEF | CAMPBELL SOUP COMPANY | 240 | ml | sodium, na | MG | 830.4 |
| SWANSON BROTH BEEF | CAMPBELL SOUP COMPANY | 240 | ml | energy | KCAL | 9.6 |
| SWANSON BROTH BEEF | CAMPBELL SOUP COMPANY | 240 | ml | carbohydrate, b… | G | 1.01 |
| SWANSON BROTH BEEF | CAMPBELL SOUP COMPANY | 240 | ml | protein | G | 1.99 |
| CAMPBELL'S SLOW KETTL… | CAMPBELL SOUP COMPANY | 440 | g | sugars, total inc… | G | 1.8 |
| CAMPBELL'S SLOW KETTL… | CAMPBELL SOUP COMPANY | 440 | g | fatty acids, total… | G | 4.49 |
| CAMPBELL'S SLOW KETTL… | CAMPBELL SOUP COMPANY | 440 | g | cholesterol | MG | 26.4 |
| CAMPBELL'S SLOW KETTL… | CAMPBELL SOUP COMPANY | 440 | g | sodium, na | MG | 1597.2 |
| CAMPBELL'S SLOW KETTL… | CAMPBELL SOUP COMPANY | 440 | g | potassium, k | MG | 286 |
| CAMPBELL'S SLOW KETTL… | CAMPBELL SOUP COMPANY | 440 | g | iron, fe | MG | 1.06 |
| CAMPBELL'S SLOW KETTL… | CAMPBELL SOUP COMPANY | 440 | g | calcium, ca | MG | 70.4 |
| CAMPBELL'S SLOW KETTL… | CAMPBELL SOUP COMPANY | 440 | g | fiber, total dietary | G | 1.76 |

Figure 3: A snippet of the "row-based" approach for the USDA Branded foods table.

The second approach is a "column-based" approach where each row contains all the data for one food. This approach encapsulates all necessary data in one row, thus allowing for more immediate results than the row-based approach. The column-based approach is shown in Figure 4. Note that all nutrients required for a food are present in a single row. One benefit is that if one knows which nutrients are to be needed by all foods, the size of the database is significantly smaller, on the level of gigabytes for larger tables. For comparison, the column-based approach for the USDA Branded foods table yielded a file of around 3.47 gigabytes, whereas the row-based approach yielded one of 6.7 gigabytes. However, a major disadvantage is that adding new nutrient values that were previously undefined involves adding and populating an entire column for all rows; this is a very costly operation.

| description | brand_owner | serving_size | serving_size_un | fiber_amoun | fiber_uni | energy_amoun | energy_uni |
|---|---|---|---|---|---|---|---|
| WESSON Vegetable Oil 1 GAL | Richardson Oilseed Products (US) … | 15 | ml | 0 | G | 130.05 | KCAL |
| SWANSON BROTH BEEF | CAMPBELL SOUP COMPANY | 240 | ml | 0 | G | 9.6 | KCAL |
| CAMPBELL'S SLOW KETTLE SO… | CAMPBELL SOUP COMPANY | 440 | g | 1.76 | G | 360.8 | KCAL |
| CAMPBELL'S SLOW KETTLE SO… | CAMPBELL SOUP COMPANY | 440 | g | 1.76 | G | 360.8 | KCAL |
| SWANSON BROTH CHICKEN | CAMPBELL SOUP COMPANY | 240 | ml | 0 | G | 9.6 | KCAL |
| CAMPBELL'S SOUP BEAN AND … | CAMPBELL SOUP COMPANY | 412 | g | 9.89 | G | 251.32 | KCAL |
| SWANSON BROTH BEEF | CAMPBELL SOUP COMPANY | 411 | g | 0 | G | 16.44 | KCAL |
| PREGO SAUCES TOMATO BASIL | CAMPBELL SOUP COMPANY | 120 | ml | 3 | G | 69.6 | KCAL |
| CAMPBELL'S SOUP TOMATO | CAMPBELL SOUP COMPANY | 120 | ml | 0.96 | G | 90 | KCAL |
| CAMPBELL'S PASTA SPAGHETTI… | CAMPBELL SOUP COMPANY | 443 | g | 5.32 | G | 385.41 | KCAL |
| BUSH'S Bourbon and Brown Suga… | Bush Brothers And Company | 130 | g | 4.55 | G | 162.5 | KCAL |
| BUSH'S Organic Garbanzo Beans … | Bush Brothers And Company | 130 | g | 4.42 | G | 136.5 | KCAL |
| BUSH'S Smokehouse Tradition Gril… | Bush Brothers And Company | 130 | g | 4.68 | G | 167.7 | KCAL |
| PEPPERIDGE FARM BREAD PA… | PEPPERIDGE FARM | 57 | g | 1.03 | G | 149.91 | KCAL |
| PEPPERIDGE FARM BREAD GA… | CAMPBELL SOUP COMPANY | 50 | g | 1 | G | 180 | KCAL |
| SWANSON BROTH BEEF | CAMPBELL SOUP COMPANY | 240 | ml | 0 | G | 9.6 | KCAL |
| PEPPERIDGE FARM COOKIES GI… | PEPPERIDGE FARM | 28 | g | 0 | G | 129.92 | KCAL |
| PEPPERIDGE FARM BREAD TUS… | PEPPERIDGE FARM | 57 | g | 1.03 | G | 140.22 | KCAL |

Figure 4: A snippet of the "column-based" approach for the USDA Branded foods table.

As the USDA databases began in a column-based form, some adjustments were needed to restructure the data into the row-based form. This consisted of creation of a table with columns to hold those nutrients that were determined to be necessary alongside other valuable information present in the column-based tables, such as the FDC index or the name of the food. Since the *Menustat* tables were already in a row-based form, no such modification was necessary.

## VI. OTHER SOURCES OF DATA

Prior to the discovery of *Menustat,* other options for collection of data with regards to restaurants were investigated. The most promising approach was to write scripts that would collect the necessary data from the internet, an approach also known as *web-scraping*. The library *Beautiful Soup* [8] was used to parse the HTML obtained by Python's *requests* library [9]. The general workflow involved making an initial query to the site that contained food data, and branching out to a restaurant, scraping the data for all food items present, and then finally moving to the next restaurant. An important point to note with this approach is that it requires the target website to have consistent HTML; if a website suddenly changes the way data is organized, the scraper could break or return invalid data.

Such a script was written to handle the website *menuwithnutrition.com.* The website has restaurants ordered alphabetically, and each restaurant has a list of the types of foods available along with the actual food items, thus proving to be structured consistently enough to be able to be scraped well [10]. A screenshot of the restaurant listing from *MenuWithNutrition* is provided in

Figure 5. Note that it is organized alphabetically. When clicking a restaurant, all the foods that the restaurant serves are displayed; one can then click each food to obtain nutrient data for the food. Approximately 205,000 foods were scraped from this site, and each food entry contains the restaurant name, food name, fat, cholesterol, sodium, carbohydrate, protein, saturated fat, trans fat, fiber, and sugar amounts, along with the units for each.

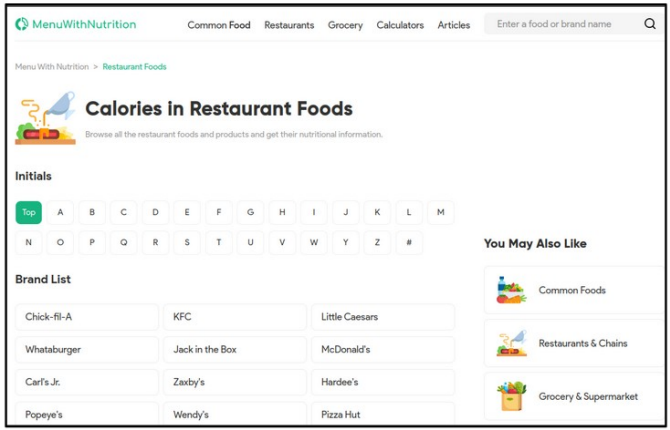

Figure 5: The restaurant listing of *MenuWithNutrition*.

## VII. FOOD IMAGES

The database can be used in combination with a folder of millions of food images. Each image has been resized so as to leave a smaller file. For databases with restaurant or brand information, the naming of the images is [RESTAURANT / BRAND NAME][FOOD NAME].png where only alphanumeric characters are saved; all non-alphanumeric characters are ignored.

In order to obtain these images, a script similar to the web-scraping script described above was created. The program used threads to split the task of querying the web and saving / resizing images among the number of cores on the client computer. It is important to note that depending on the user's internet speed and number of cores, some adjustments may need to be made to ensure that no IP blocking occurs due to the frequency of requests. This can be done by altering the amount each processor sleeps.

## VIII. DATA REPOSITORY:

All database files and the respective code that generated them can be found at:

https://github.com/lxaw/ComprehensiveFoodDatabase

## IX. CONCLUSION

While many studies make use of food data, there is a lack of reliable and free sources of this information. This project investigated the merits of different data sources and table designs, and explained how one could create tables from these sources or scrape images. The goal of this project was to provide a potential solution, a database that had a suitable number of entries for both restaurant and non-restaurant foods, while also providing images for each entry. However, the main goal is to leave a usable free database for future researchers to allow for greater research to be done at less cost.

## X. ACKNOWLEDGMENT

Research reported in this publication was supported by grants from the National Institutes of Health.

## XI. VIII. REFERENCES